\documentstyle[prl,aps,multicol,psfig,array]{revtex}

\newcommand{\beq}{\begin{equation}}
\newcommand{\eeq}{\end{equation}}
\newcommand{\bqa}{\begin{eqnarray}}
\newcommand{\eqa}{\end{eqnarray}}

\draft

\begin{document}

\author{J. O. Andersen, U Al Khawaja and H. T. C. Stoof}
\address{\it Institute for Theoretical Physics,
             Utrecht University,
             Leuvenlaan 4, 3584 CE Utrecht, The Netherlands\\
             {\rm (Received \today)}}

\title{Phase fluctuations in atomic Bose gases}

\maketitle

\begin{abstract}
We improve on the Popov theory for partially Bose-Einstein
condensed atomic gases by treating the phase fluctuations
exactly. As a result, the theory becomes valid in arbitrary
dimensions and is able to describe the low-temperature crossover
between three, two and one-dimensional Bose gases, which is
currently being explored experimentally. 
We consider both homogeneous and trapped Bose gases.
\end{abstract}

\pacs{PACS numbers: 03.75.Fi, 67.40.-w, 32.80.Pj}

\begin{multicols}{2}

{\it Introduction.} --- One of the most important features of the
trapped Bose-Einstein condensed atomic gases is the remarkable control
that can be achieved experimentally over the relevant physical
parameters. As a result these gases can
be studied in various physically distinct regimes, where
their properties are quite different. The latest achievement in
this respect is the experiment by G\"orlitz {\it et
al.}~\cite{lowdketterle}, in which one and two-dimensional
Bose-Einstein condensates were created by reducing the temperature
and the average interaction energy of the atoms below the energy
splitting of either two or one of the directions of the harmonic
trapping potential, respectively.

As is well known, the physics of one and two-dimensional systems
is fundamentally different from the physics in three dimensions.
This difference is for instance illustrated by the famous
Mermin-Wagner-Hohenberg theorem~\cite{mermin,hohen}, which states
that in one dimension Bose-Einstein condensation in a
homogeneous Bose gas never occurs, whereas in two dimensions 
a condensate
can only exist at zero temperature. In both cases, this is due to
the enhanced importance of phase fluctuations. The
Mermin-Wagner-Hohenberg theorem is valid only in the thermodynamic
limit and does not apply to finite-size systems. In one or
two-dimensional trapped Bose gases a condensate can therefore
exist if the external trapping potential sufficiently restricts
the size of the atomic gas cloud~\cite{mullin,jason}.

The physics of low-dimensional Bose gases is in fact even more
interesting, because, notwithstanding the
Mermin-Wagner-Hohenberg theorem, a dilute homogeneous
two-dimensional Bose gas is expected to undergo at a nonzero
critical temperature a true thermodynamic phase transition, that
is known as the Kosterlitz-Thouless transition~\cite{koster}.
Below the critical temperature, the gas is superfluid but has only
algebraic long-range order. This so-called topological phase
transition is therefore not characterized by a local order
parameter, but by the unbinding of vortex pairs and the resulting
destruction of superfluidity. Since the Mermin-Wagner-Hohenberg
theorem forbids a true Bose-Einstein condensate in two dimensions,
the superfluid phase is only characterized by the existence of a
so-called ``quasicondensate''. This important concept was first
introduced by Popov \cite{popov} and roughly speaking corresponds
to a condensate with a fluctuating phase.

Although introduced theoretically as early as in 1972, the actual
observation of such a quasicondensate has only very recently been
made in a spin-polarized atomic hydrogen adsorbed on a superfluid
$^4$He surface by Safonov {\it et al.}~\cite{saf}. In particular,
this experiment measures the three-body (dipolar) recombination
rate of a spin-polarized atomic hydrogen gas and observes a
reduction in the associated rate constant due to the presence of a
quasicondensate. Qualitatively, this reduction was anticipated by
Kagan, Svistunov and Shlyapnikov, but the magnitude of the effect
turns out to be much larger than predicted \cite{kagan}. The reason
for this discrepancy is still not fully understood,
although a physical mechanism for the additional reduction was
already suggested by Stoof and Bijlsma before its observation
\cite{henk2}.

In the context of trapped atomic Bose gases, the possibility of
observing a quasicondensate has been explored by Petrov {\it et
al.}~\cite{1d,2d}. Although this presents an important first step
towards understanding the physics of low-dimensional Bose gases,
the approach can only be justified close to zero
temperature. The reason for this is that density fluctuations are
neglected from the outset. As a result, depletion of the
quasicondensate, due to either quantum or thermal fluctuations,
cannot be properly accounted for. In particular, the approach
does not lead to an equation of state for the Bose gas.

The main aim of the present Letter is to
overcome this problem and to formulate a microscopic theory that
includes both density and phase fluctuations of the Bose gas. It
is similar in spirit to the successful Popov theory for
three-dimensional Bose-Einstein condensed gases, and can be used
to study in detail the thermodynamic behavior of low-dimensional
degenerate Bose gases. Moreover, it describes also the crossover
between three, two and one-dimensional Bose gases. 
To explain most clearly how this can be achieved, we first discuss the
homogeneous case.  After that, we
generalize the theory to inhomogeneous gases.

{\it Modified Popov theory.} --- Formulating a microscopic theory
for lower-dimensional 
Bose gases is complicated by the fact
that mean-field theory is plagued with infrared divergences. To
facilitate the discussion of how to deal with these divergences,
we first recapitulate the expressions for the density $n$ and the
chemical potential $\mu$ that follow from the usual Popov theory
for partially Bose-Einstein condensed gases. For a Bose gas in a
box with volume $V$ they read \cite{popov,henk1}
\begin{eqnarray}
\label{1ln}
n=n_0 &+& {1\over V}\sum_{\bf k} \Bigg[ {\epsilon_{\bf
k}+n_0V_0-\hbar\omega_{\bf k}\over2\hbar\omega_{\bf k}} \nonumber
\\ && \hspace*{0.5in} + {\epsilon_{\bf k}+n_0V_0\over\hbar\omega_{\bf k}}
N(\hbar\omega_{\bf k}) \Bigg]\;, \\
\label{1lmu}
\frac{\mu}{V_0} = n_0 &+& {1\over V} \sum_{\bf k} \Bigg[ {2\epsilon_{\bf
k}+n_0V_0-2\hbar\omega_{\bf k}\over2\hbar\omega_{\bf k}} \nonumber
\\&& \hspace*{0.5in} + {2\epsilon_{\bf k}+n_0V_0\over\hbar\omega_{\bf k}}
N(\hbar\omega_{\bf k})\Bigg]\;.
\end{eqnarray}
Here $n_0$ is the density of the Bose-Einstein condensate,
$V_0 \delta({\bf x}-{\bf x}')$ is the
bare two-body interaction potential, $\epsilon_{\bf k} = \hbar^2{\bf k}^2/2m$
is the kinetic energy of the atoms, $\hbar\omega_{\bf k}=(\epsilon_{\bf
k}^2+2n_0V_0\epsilon_{\bf k})^{1/2}$ is the Bogoliubov dispersion
relation, and $N(x)=1/(e^{\beta x}-1)$ is the Bose-Einstein distribution
function where $ \beta = 1/k_{\rm B} T$ is the inverse thermal energy.

In agreement with the Mermin-Wagner-Hohenberg theorem, the
momentum sums in Eqs.~(\ref{1ln}) and (\ref{1lmu}) contain terms
that are infrared divergent at all temperatures in one dimension
and at any nonzero
temperature in two dimensions. The physical
reason for these ``dangerous'' terms is that the above expressions
have been derived by taking into account only quadratic
fluctuations around the classical result $n_0$, i.e., by writing
the annihilation operator for the atoms as $\hat{\psi}({\bf x}) =
\sqrt{n_0} + \hat{\psi}'({\bf x})$ and neglecting in the
hamiltonian terms of third and fourth order in $\hat{\psi}'({\bf
x})$. As a result the phase fluctuations of the condensate give
the quadratic contribution $n_0 \langle \hat{\chi}({\bf x})
\hat{\chi}({\bf x}) \rangle$ to the right-hand side of the above
equations, whereas an exact approach that sums up all the
higher-order terms in the expansion would clearly give no
contribution at all to these local quantities because $\langle
e^{-i\hat{\chi}({\bf x})} e^{i\hat{\chi}({\bf x})} \rangle = 1 +
\langle \hat{\chi}({\bf x}) \hat{\chi}({\bf x}) \rangle + \dots =
1$. To correct for this we thus need to subtract the quadratic
contribution of the phase fluctuations, which from
Eqs.~(\ref{1ln}) and (\ref{1lmu}) is seen to be given by
\begin{equation}
\label{phase}
n_0 \langle \hat{\chi}({\bf x}) \hat{\chi}({\bf x}) \rangle =
{1\over V} \sum_{\bf k} {n_0 V_0\over2\hbar\omega_{\bf k}}
\left[1+2N(\hbar\omega_{\bf k})\right]~.
\end{equation}
As expected, the infrared divergences that occur in the one and
two-dimensional case are removed by performing this subtraction.

After having removed the spurious contributions from the phase
fluctuations of the condensate, the resulting expressions turn out
to be ultraviolet divergent. These divergences can be removed by
the standard renormalization of the bare coupling constant $V_0$.
Apart from a subtraction, this essentially amounts to replacing
everywhere the bare two-body potential $V_0$ by the two-body
T-matrix evaluated at zero initial and final relative momenta and
at the energy $-2\mu$, which we denote from now on by $T^{\rm
2B}(-2\mu)$. Note that the energy argument of the T-matrix is
$-2\mu$, because this is precisely the energy it costs to excite two
atoms from the condensate \cite{schick,fisher}. In this manner, we
finally arrive at
\bqa
\label{h1}
n&=&n_0+{1\over V}\sum_{\bf k}
\Bigg[
{\epsilon_{\bf k}-\hbar\omega_{\bf k}\over2\hbar\omega_{\bf k}}
+{n_0T^{\rm 2B}(-2\mu)\over2\epsilon_{\bf k}+2\mu} \nonumber \\
&&\hspace*{0.8in} 
+
{\epsilon_{\bf k}\over\hbar\omega_{\bf k}}
N(\hbar\omega_{\bf k})
\Bigg]\;, \\
\mu&=&(2n-n_0)T^{\rm 2B}(-2\mu) 
\label{h2} 
=(2n'+n_0)T^{\rm 2B}(-2\mu) \;,
\eqa
where $n'=n-n_0$ represents the depletion of the condensate
due to quantum and thermal fluctuations and
the Bogoliubov quasiparticle dispersion now equals
$\hbar\omega_{\bf k}
=\left[\epsilon_{\bf k}^2+2n_0T^{\rm 2B}(-2\mu)\epsilon_{\bf k}\right]^{1/2}$.
The most important feature of Eqs.~(\ref{h1}) and (\ref{h2}) is that
they contain no infrared and ultraviolet divergences and therefore
can be applied in any dimension and at all temperatures, even if
no condensate exists. How this can be reconciled with the
Mermin-Wagner-Hohenberg theorem is discussed next.

{\it One dimension.} --- To understand the physical meaning of the
quantity $n_0$ in Eqs.~(\ref{h1}) and (\ref{h2}), we must determine
the off-diagonal long-range behavior of the one-particle density
matrix. Because this is a nonlocal property of the Bose gas, the
phase fluctuations contribute and 
in the limit $|{\bf x}|\rightarrow\infty$, we find 
\begin{equation}
\langle\hat{\psi}^{\dagger}({\bf x}) \hat{\psi}({\bf 0})\rangle
\simeq
n_0 e^{-\langle \left[ \hat{\chi}({\bf x})-\hat{\chi}({\bf 0})
    \right]^2 \rangle/2}\;.
\end{equation}
Moreover, using 
Eq.~(\ref{phase})
and carrying out the renormalization
of the bare coupling constant, we obtain 
\begin{eqnarray}\nonumber
\langle \left[\hat{\chi}({\bf x}) -
             \hat{\chi}({\bf 0}) \right]^2 \rangle
            &=& {T^{\rm 2B}(-2\mu)\over V} \sum_{\bf k}
               \left[{1\over\hbar\omega_{\bf k}}
\left[1+2N(\hbar\omega_{\bf k})\right]
\right.
\\ &&\left.
-{1\over\epsilon_{\bf k}+\mu} \right]
        \left[1-\cos({\bf k}\!\cdot\!{\bf x})\right]\;.
\label{pp}
\end{eqnarray}
At zero temperature, the quantity $\langle \left[\hat{\chi}({\bf
x}) - \hat{\chi}({\bf 0}) \right]^2 \rangle$ diverges
logarithmically for large distances, which leads to algebraic
off-diagonal long-range order in the one-particle density matrix. In
detail we can show that the leading behavior of the
zero-temperature one-particle density matrix for $|{\bf x}|\gg
\xi$ is
\begin{equation}
\label{exact} \langle\hat{\psi}^{\dagger}({\bf x}) \hat{\psi}({\bf
0})\rangle \simeq {n_0\over(|{\bf x}|/\xi)^{\eta}}\;,
\end{equation}
where $\eta = 1/4\pi n_0\xi$ is the correlation-function exponent
and  $\xi = \hbar/[4mn_0T^{\rm 2B}(-2\mu)]^{1/2}$ is the correlation
length. At nonzero temperatures, $\langle \left[\hat{\chi}({\bf x})
- \hat{\chi}({\bf 0}) \right]^2 \rangle$ diverges linearly for
large distances and the one-particle density matrix thus no longer
displays off-diagonal long-range order.

A few remarks are in order at this point. Most importantly for
our purposes, the asymptotic behavior of the one-particle density
matrix at zero temperature proves that the gas is not
Bose-Einstein condensed and that $n_0$ should be
identified with the quasicondensate density. Moreover, from our
equation of state we can show that in the weakly-interacting limit
$4\pi n \xi \gg 1$, the fractional depletion of the
quasicondensate is $(n-n_0)/n=(\pi\sqrt{2}/4-1)/4\pi n \xi$
and therefore very small. Keeping this in mind, Eq.~(\ref{exact})
is in complete agreement with the exact result obtained by
Haldane~\cite{haldane}. Note that our theory cannot 
describe the strongly-interacting case $4\pi n \xi \ll 1$, where
the one-dimensional Bose gas behaves as a 
Tonks gas
\cite{olshanii,GW}. Finally, our results show that at nonzero
temperatures not even a quasicondensate exists and we have to use
the equation of state for the normal state
$n=\sum_{\bf k} N(\epsilon_{\bf k} + \hbar\Sigma - \mu)/V$
to describe the gas. Here, the Hartree-Fock self-energy satisfies
$\hbar\Sigma=2nT^{\rm 2B}(-\hbar\Sigma)$.

{\it Two dimensions.} --- Applying the same arguments in two
dimensions leads to the conclusion that at zero temperature, $n_0$
corresponds to the condensate density, whereas at a nonzero
temperature, it represents the quasicondensate density. In
particular, the correlation-function exponent is $\eta =
1/n_0\Lambda^2$ where $\Lambda = (2\pi\hbar^2/mk_{\rm B}T)^{1/2}$
is the thermal de Broglie wave length. Due to the mean-field nature
of the modified Popov theory, the Kosterlitz-Thouless transition
is absent and a nontrivial solution of the equation of state
exists even if $\eta > 1/4$. This can be corrected for by
explicitly including the effect of vortex pairs in the phase
fluctuations. As we show in a future paper, this is achieved by
using the modified Popov theory to determine the initial values of
a renormalization-group calculation for the superfluid density and
the fugacity of the vortices. It should, however, be noted that
for many applications we are not interested in the phase
fluctuations 
and this additional renormalization is not very important.
This is for instance true for the reduction of the three-body 
recombination rate constant of a hydrogen gas~\cite{future}. 

At zero temperature, the fractional depletion of the condensate in
the Popov approximation was first calculated by
Schick~\cite{schick} and is $T^{\rm 2B}(-2\mu)/4\pi$ where the
chemical potential satisfies $\mu = n T^{\rm 2B}(-2\mu)$. The
corresponding result based on Eq.~(\ref{h1}) is
$\left(1-\ln2\right)T^{\rm 2B}(-2\mu)/4\pi$ where $\mu$ now satisfies
Eq.~(\ref{h2}). Thus the depletion is reduced by a
factor of approximately 3.
In two dimensions, the two-body T-matrix 
cannot be approximated by an energy-independent constant 
and this is the reason why we have been very careful about the appropriate
energy of the two-body collisions~\cite{com}. The T-matrix 
now obeys
\bqa
T^{\rm 2B}(-2\mu)&=&{4\pi\hbar^2\over m}
{1\over\ln(2\hbar^2/\mu ma^2)}\;,
\eqa
where $a$ is the scattering length.

{\it Three dimensions.} --- We know that the Popov theory has been
very successful in describing the properties of three-dimensional
trapped Bose gases. It is therefore important to mention that,
although the modification that we have performed is essential for
one and two-dimensional Bose gases, it leads only to minor changes
in the three-dimensional case. This can be seen by
considering the temperature dependence of the condensate density.
At zero temperature, the fractional depletion that results from the
Popov theory was first calculated by Lee and Yang~\cite{leeyang}
and equals $(8/3)\sqrt{na^3/\pi}$, where the T-matrix is taken
to be $T^{\rm 2B}(-2\mu)=4\pi a\hbar^2/m$ and $a$ is the 
scattering length. 
The result that follows from
Eqs.~(\ref{h1}) and~(\ref{h2}) is
$\left({32/3}-2\sqrt{2}\pi\right)\sqrt{na^3/\pi}$. 
The fractional depletion is thus approximately 2/3
of the Popov result.
This is in fact the largest change in the condensate depletion,
since the effects of the phase fluctuations decreases at larger temperatures.
The critical temperature $T_{\rm BEC}$ is found by taking the limit
$n_0\rightarrow 0$ in Eqs.~(\ref{h1}) and~(\ref{h2}).
These expressions then reduce to the 
same expressions for the density and chemical
potential as in the Popov theory.
This implies that our critical temperature for Bose-Einstein
condensation coincides with  
that obtained in the Popov theory and 
with that of an ideal Bose gas.

{\it Trapped Bose gases.} ---
We now generalize our mean-field theory to a Bose gas in an external potential
$V^{\rm ext}({\bf x})$. First, the right-hand side of Eq.~(\ref{h1})
is expressed in terms of the Bogoliubov coherence factors 
$u_{\bf k}$ and $v_{\bf k}$, which are then replaced by 
$u_j({\bf x})$ and $v_j(\bf x)$. The latter satisfy 
the Bogoliubov-de Gennes equations~\cite{pita,fetter} 
\bqa\nonumber
\hspace{-0.1cm}
\Bigg[\!
-\hbar\omega_j\! 
-{\hbar^2\over 2m}\nabla^2\!
-\mu({\bf x})\!
+2T^{\rm 2B}(-2\mu({\bf x})) 
n({\bf x})\Bigg]u_j({\bf x})
&&
\nonumber \\ 
\hspace{-2.5cm}
+T^{\rm 2B}(-2\mu({\bf x}))
n_0({\bf x})v_j({\bf x})
=0\;,&&
\label{u} \\ \nonumber
\Bigg[
\hbar\omega_j 
\!-{\hbar^2\over 2m}\nabla^2
\!-\mu({\bf x})+2T^{\rm 2B}(-2\mu({\bf x}))
n({\bf x})\Bigg]v_j({\bf x})
&&\\  
+T^{\rm 2B}(-2\mu({\bf x}))
n_0({\bf x})u_j({\bf x})=0\;,
\label{v}
\eqa
where $n_0({\bf x})=|\psi_0({\bf x})|^2$ and 
$\mu({\bf x})=\mu-V^{\rm ext}({\bf x})$.
Note that we have
chosen $u_j({\bf x})$ and $v_j({\bf x})$ to be real. In some cases, 
e.g. when the macroscopic wavefunction
$\psi_0({\bf x})$ contains a vortex, one cannot choose
these amplitudes to be real and our equations must be generalized to 
incorparate that fact.
In terms of these particle and hole
amplitudes, the expression for the total density in 
Eq.~(\ref{h1}) reads
\bqa\nonumber
n({\bf x})&=&n_0({\bf x})+
\sum_j
\Bigg\{\left[
u_j({\bf x})+v_j({\bf x})\right]^2N(\hbar\omega_j)
+v_j({\bf x})
\\ &&
\hspace{-1cm}
\times[u_j({\bf x})+v_j({\bf x})]
+{T^{\rm 2B}(-2\mu({\bf x}))n_0({\bf x})
\over 2\epsilon_j+2\mu({\bf x})}[\phi_j({\bf x})]^2
\Bigg\}\;.
\label{ntotal}
\eqa
Here, $\epsilon_j$ are the eigenenergies 
and $\phi_j({\bf x})$  the eigenstates 
of the external trapping potential.
In the large-$j$ limit, we have that
$u_j({\bf x})=\phi_j({\bf x})$ and 
\begin{equation}
v_j({\bf x})=-{T^{\rm 2B}(-2\mu({\bf x}))n_0({\bf x})
\over2\epsilon_j}\phi_j({\bf x})\;.
\label{vv}
\end{equation}   
As a result, the sum in the expression for the total density is
ultraviolet finite 
since the second and third term cancel each other in the large-$j$ limit.
Furthermore, the inhomogeneous generalization of Eq.~(\ref{h2}) 
becomes the nonlinear Schr\"odinger equation
\bqa\nonumber
\mu\psi_0({\bf x})&=&
\Bigg[
-{\hbar^2\over 2m}\nabla^2
+V^{\rm ext}({\bf x})
\\ && \hspace{-1cm}
+T^{\rm 2B}(-2\mu({\bf x}))
(2n^{\prime}({\bf x})+|\psi_0({\bf x})|^2)
\Bigg]\psi_0({\bf x})\;.
\label{gpu}
\eqa

Finally, the expression for the phase fluctuations
in the case of a trapped Bose gas can be obtained
from 
$\langle\hat{\chi}({\bf x})\hat{\chi}({\bf x}^{\prime})\rangle$,
which is equal to the part of
\bqa
\nonumber
-\sum_j
{1\over\sqrt{n_0({\bf x})n_0({\bf x}^{\prime})}}
\Bigg\{u_j({\bf x})v_j({\bf x}^{\prime})\Big[1+2N(\hbar\omega_j)\Big]
&&\\ 
+\Bigg[{T^{\rm 2B}(-2\mu({\bf x}^{\prime}))n_0({\bf x}^{\prime})
\over2\epsilon_j+2\mu({\bf x}^{\prime})}
\Bigg]
\phi_j({\bf x})\phi_j({\bf x}^{\prime})
\Bigg\}
&&\;,
\label{chi}
\end{eqnarray}
which is symmetric under exchange of ${\bf x}$  and ${\bf x}^{\prime}$.
Note that this correlation function is free of ultraviolet divergences
in contrast to the long-wavelength result used in Refs.~\cite{1d,2d}.

{\it Discussion.} --- We have proposed a new mean-field theory for
dilute Bose gases in arbitrary dimensions, in which the phase
fluctuations are treated exactly. 
We reproduce exact results in
one dimension and the results in three dimensions 
are essentially the same as those predicted by the Popov theory. 
The exact treatment of the phase fluctuations has solved the long-standing
problem of infrared divergences in one and two dimensional Bose systems.
This opens up the possibility to study in the same
detail as in the three-dimensional
case, the physics of low-dimensional atomic gases.
As previously mentioned, we can also incorporate the Kosterlitz-Thouless
transition and for instance perform an {\it ab initio}
calculation of the critical temperature as 
a function of the scattering length and the 
density of the gas.

The inhomogeneous generalization of the modified Popov theory
describes in a selfconsistent manner the
density profile 
and the full crossover from a condensate to quasicondensate 
in one and two-dimensional trapped Bose gases.
In practice, however, we expect
that a good approximation 
is obtained by simply calculating 
the densities $n_0({\bf x})$ and $n'({\bf x})$ in the Thomas-Fermi
approximation, i.e., by applying Eqs.~(\ref{h1}) and (\ref{h2})
locally at every point in space
with a chemical potential equal to $\mu({\bf x})$. 
Work in this direction is also in
progress~\cite{future}.\\

We thank Tom Bergeman 
for valuable discussions. 
This work was
supported by the Stichting voor Fundamenteel Onderzoek der Materie
(FOM), which is supported by the Nederlandse Organisatie voor Wetenschapplijk 
Onderzoek (NWO).


\end{multicols}

\begin{thebibliography}{99}
\bibitem{com} Note that the appropriate energy for a collision between
a condensate and thermal atom is $-\mu$ instead of $-2\mu$. For simplicity,
we have not made that distinction here.
\end{thebibliography}
\end{document}